\documentclass[aps,pra]{revtex4}
\usepackage[english]{babel}
\usepackage{amsmath}
\usepackage{epsfig}
\usepackage{dcolumn}

\begin{document}
%\draft

\title{VIRIAL THEOREM AND GIBBS THERMODYNAMIC POTENTIAL FOR COULOMB SYSTEMS}

\author{V.B. Bobrov $^{1,2}$, S. A. Trigger $^{1,\,3}$}
\address {$^1$ Joint Institute for High Temperatures, Russian Academy of
Sciences, Izhorskaya St., 13, Bd. 2, 125412 Moscow, Russia;\\
$^2$ National Research University "MPEI"\,,
Krasnokazarmennaya str. 14, Moscow, 111250, Russia;\\
$^3$ Institut f\"ur Physik, Humboldt-Universit\"at zu Berlin,
Newtonstra{\ss}e 15, D-12489 Berlin, Germany;\\
 e-mails: vic5907@mail.ru,\;satron@mail.ru\\}

\begin{abstract}
Using the grand canonical ensemble and the virial theorem, we show that the Gibbs thermodynamic potential of the non-relativistic system of charged particles is uniquely defined by single-particle Green functions of electrons and nuclei. This result is valid beyond the perturbation theory with respect to the interparticle interaction. \\

PACS number(s): 05.30.Fk, 05.70.Ce, 51.30.+i, 52.27.Gr

\end{abstract}

\maketitle
In a wide range of thermodynamic parameters, matter properties are most adequately described in the model of the non-relativistic system of charged particles (electrons and nuclei) interacting via the Coulomb potential (Coulomb system, CS) [1]. Difficulties in the theoretical study of CS properties are caused by the necessity to simultaneously consider collective effects, first of all, Coulomb interaction screening, and the quantum description, either due to the formation of bound states of electrons and nuclei, and the necessity to consider the identity of elementary particles [2]. The principal possibility of solving the arisen problems is based on various versions of the diagram technique of the perturbation theory with respect to the interparticle interaction, developed within quantum field theory methods (see, e.g., [3, 4]).
When considering the thermodynamic properties of CSs in the volume $V$ at temperature $T$ (in energy units), the problem is reduced to the calculation of the Gibbs thermodynamic potential $\Omega(V,T, {\mu_a})$
\begin{eqnarray}
\Omega(V,T, {\mu_a})=-T \ln \left\{\mathrm{Tr}\exp \left(-\frac{\hat{H}-\sum_b \mu_b \hat{N}_b}{T}\right)\right\}, \label{F1}
\end{eqnarray}
where the CS Hamiltonian is given by
\begin{eqnarray}
\hat{H}=\hat{H}^{(0)}+\hat{U}^{int},\;  \hat{H}^{(0)}=\sum_a \sum_{{\bf p},\sigma}\epsilon_a (p) \hat{a}_{{\bf p}\sigma}^{+}\hat{a}_{{\bf p}\sigma}, \; \hat{U}^{int}=\sum_{a,b} \hat{U}_{ab},\; \nonumber\\ \hat{U}_{ab}=\frac{1}{2V}\sum_{{\bf q}\neq 0} \sum_{\{{\bf p}\sigma \}} u_{ab}({\bf q}) \hat{a}_{{\bf p}_1+{\bf q}/2,\sigma_1}^{+}\hat{b}_{{\bf p}_2-{\bf q}/2,\sigma_2}^{+} \hat{b}_{{\bf p}_2+ {\bf q}/2,\sigma_2} \hat{a}_{{\bf p}_1-{\bf q}/2,\sigma_1 }.
\label{F2}
\end{eqnarray}
Here $\hat{a}^{+}_{{\bf p},\sigma}$ and $\hat{a}_{{\bf p},\sigma}$ are the creation and annihilation operators, respectively, with the momentum $\hbar{\bf p}$ and spin number $\sigma$ for particles of type $a$, which are characterized by the mass $m$, charge $z_a e$, and chemical potential $\mu_a$, $\hat{N_a} = \sum_{{\bf p},\sigma}\hat{a}^{+}_{{\bf p},\sigma}\hat{a}_{{\bf p},\sigma}$ is the operator of the total number of particles of type $a$, $\epsilon_a(p)=\frac{\hbar^2 p^2}{2 m_a}$ is the energy of free particles of type $a$, and $u_{ab}(q)=4\pi z_a z_b e^2/q^2$ is the Fourier component of the Coulomb interaction potential of particles of types $a$ and $b$. In this case, the chemical potentials $\mu_a$ of various particle types are considered as formally independent if the CS quasi-neutrality conditions
\begin{eqnarray}
\sum_{a}z_a e n_a=0,
\label{F3}
\end{eqnarray}
are taken into account at the final stage of calculations of physical quantities (see [5] for more details). Here $n_a (T, \mu_b)=\langle\hat{N_a}\rangle/V=-(\partial \Omega/\partial \mu_a)_{V,T, \mu_b\neq\mu_a}$ is the average density of the number of particles of type $a$ in the volume $V$, angle brackets $\langle...\rangle$ mean averaging with the grand canonical ensemble.

The Gibbs thermodynamic potential $\Omega (V,T,\mu_a)$ (1) is calculated in the theory of equilibrium CSs using three approaches based on diagram techniques of the perturbation theory. One of these approaches is based on the Matsubara diagram technique [6]. A detailed description of this approach for determining the thermodynamic quantities of hydrogen gas plasma at the present stage is given in [7]. In the other approach, the Kadanoff and Baym diagram technique [3] is used; it is based on the consideration of the time-dependent Green functions and allowing uniform consideration of both equilibrium and nonequilibrium CS properties (see, e.g., [8]). The state of the art of the approach based on the Kadanoff and Baym diagram technique is presented in [5, 9]. An alternative approach to the consideration of CS thermodynamic properties was proposed by Montroll and Ward [10]. They generalized the Mayer method of group integrals and the corresponding diagram technique to the quantum case. The development of this method based on screened cluster expansion (SCE) within loop formalism is presented in [11]. In contrast to two other approaches based on the Green function formalism, the rules for calculating diagrams within the SCE are very complex. This factor limits the possibility of an analytical study using the simplest approximations. However, a significant advantage of this approach is the possibility of the direct application of the path integral Monte Carlo (PIMC) method to calculate CS thermodynamic properties beyond the perturbation theory (see [11--13] and references therein). Thus, the Montroll and Ward approach can be used to study CS equilibrium properties by numerical methods and to compare the results obtained in such a way with analytical results of two other approaches based on the Green function formalism. In this case, we proceed from the fact that exact consideration shows that all three approaches stated above are equivalent.

However, particular analytical calculations are restricted to the consideration of only a limited number of expansion terms in series of the perturbation theory. In this case, equivalent results of the considered approaches appear only when considering the simplest approximations corresponding to the lowest orders of the perturbation theory. In particular, a unique solution to the known problem of the calculation of the finite statistical sum of the hydrogen "atom" in rarefied plasma has not yet been obtained (see [14, 15] and references therein).

In this case, exact relations for the Gibbs thermodynamic potential of the CS becomes especially important. To this end, we use the virial theorem; for the CS, it is written as
\begin{eqnarray}
P V=\frac{2}{3}\langle\hat{H}^{(0)}\rangle+\frac{1}{3}\langle\hat{U}^{int}\rangle,
\label{F4}
\end{eqnarray}
where $P$ is the pressure in the CS. Let us pay attention that the virial theorem is valid for describing equilibrium systems of any Gibbs distribution (see, e.g., [16]). This circumstance is a consequence of the fact that the virial theorem can be obtained directly from the stationary Schrodinger equation [17, 18]. This means that, in relation (5), we can consider the pressure as $P=P(T,{\mu_b})$ or, that is equivalent, $P(T,{\mu_b})V=-\Omega(V, T,{\mu_b})$. Thus, the virial theorem can be used as an alternative method for calculating the Gibbs thermodynamic potential, hence, any thermodynamic properties of the CS.

Now take into account that the quantity $\langle\hat{H}^{(0)}\rangle$ can be written as
\begin{eqnarray}
\langle\hat{H}^{(0)}\rangle=V \sum_a\sigma \int \frac{d^3 p}{(2\pi)^3} \epsilon_a(p)f_a (p,\sigma); \qquad f_a (p,\sigma)=\langle\hat{a}^{+}_{{\bf p},\sigma}  \hat{a}_{{\bf p},\sigma}\rangle,
\label{F5}
\end{eqnarray}
where $f_a(p,\sigma)$ is the one-particle distribution function (average occupation number) for particles of type $a$, which also depends on the temperature $T$ and various chemical potentials $\mu_b$. For the system of noninteracting particles, the function $f_a(p,\sigma)$ is defined by the Fermi-Dirac or Bose-Einstein distributions. For the average density $n_a$, we can use the equality [2]
\begin{eqnarray}
n_a (T,{\mu_b})=\sum_\sigma \int \frac{d^3 p}{(2\pi)^3} f_a (p,\sigma),
\label{F6}
\end{eqnarray}
In the Kadanoff and Baym technique, the one-particle distribution function is directly related to the one-particle correlation function $g^{<}_a(p,\sigma;\omega)$ (see [2, 3] for more details)
\begin{eqnarray}
f_a (p,\sigma)= \int_{-\infty}^\infty \frac{d \omega}{2\pi} g^{<}_a(p,\sigma;\omega).
\label{F7}
\end{eqnarray}
In turn, the function $g^{<}_a(p,\sigma;\omega)$ uniquely defines the corresponding single-particle Green function (SPGF) taking into account all interaction effects and quantum statistics [2, 3]. Then we should define the method for calculating the average potential energy $\langle\hat{U}^{int}\rangle$ in the CS. Within the Kadanoff and Baym approach [3], we have
\begin{eqnarray}
\langle\hat{U}^{int}\rangle= \frac{V}{2}\sum_a \sum_\sigma  \int \frac{d^3 p}{(2\pi)^3} \int_{-\infty}^\infty \frac{d \omega}{2\pi}\,[\omega-\varepsilon_a(p)]\, g^{<}_a(p,\sigma;\omega).
\label{F8}
\end{eqnarray}
Thus, taking into account the virial theorem (5), we come to the conclusion that the Gibbs thermodynamic potential for the CS is uniquely defined by the SPGF of electrons and nuclei beyond the perturbation theory using the relation
\begin{eqnarray}
P(T,{\mu_b})= \frac{1}{6}\sum_a \sum_\sigma  \int \frac{d^3 p}{(2\pi)^3} \int_{-\infty}^\infty \frac{d \omega}{2\pi}\,[\omega+3\varepsilon_a(p)]\, g^{<}_a(p,\sigma;\omega).
\label{F9}
\end{eqnarray}
This statement is consistent with the known result by Luttinger and Ward [19], i.e., the Gibbs thermodynamic potential for a single-component system is an SPGF functional. The explicit form of this functional can be determined only within the perturbation theory, although it is attempted to solve this problem within thermodynamically self-consistent approximations (see [20, 21] and references therein). 

The remarkable peculiarity of the Coulomb system is the possibility to obtain the explicit representation of the thermodynamic potential as the functional of the Green functions for electrons and positive charges. In contrast with the case of the short-range potential, this representation is not an infinite series on the respective Green functions, but has the simple form (9).
Therefore, the problem of the calculation of CS thermodynamic properties is reduced to the determination of the SPGFs of electrons and nuclei only.\\

This study was supported by joint grant of Russian Foundation for Basic Research and Ukraine National Academy of Science, projects No. 12-08-00822-a and No. 12-02-90433-Ukr-a.

\end{document}